\documentclass[a4paper,useAMS,usenatbib]{mnras}
\usepackage{graphics,epsfig,psfig}
\usepackage[normalem]{ulem}
\usepackage{xcolor}
\usepackage{float}
\usepackage{subfig}
\usepackage{graphicx}
\usepackage[]{inputenc,amssymb}

\usepackage[style=base, singlelinecheck=off, font=small, labelfont=bf, justification=raggedright]{caption}
\def \be{\begin{equation}}
\def \ee{\end{equation}}
\def \bea{\begin{eqnarray}}
\def \eea{\end{eqnarray}}

\def\apj{ApJ}
\def\mnras{MNRAS}
\def\prd{PRD}
\def\aap{A\&A}
\definecolor{webgreen}{rgb}{0,.5,0}
\definecolor{webbrown}{rgb}{.6,0,0}

\title[Reionization as obstacle to primordial gravitational waves]{Is patchy reionization an obstacle in detecting the primordial gravitational wave signal?}
\author[Mukherjee, Paul \& Choudhury]{Suvodip Mukherjee$^{1,2,3}$\thanks{mukherje@iap.fr}, Sourabh Paul$^{4,5}$ \thanks{sourabh.paul@gmail.com} \& Tirthankar Roy Choudhury$^5$ \thanks{tirth@ncra.tifr.res.in}\\
$^{1}$ Institut d'Astrophysique de Paris,  98bis Boulevard Arago, 75014 Paris, France\\
$^{2}$ Sorbonne Universites, Institut Lagrange de Paris,  98 bis Boulevard Arago, 75014 Paris, France\\
$^{3}$ Center for Computational Astrophysics, Flatiron Institute, 162 5th Avenue, 10010, New York, NY, USA\\
$^{4}$ Department of Physics and Astronomy, University of the Western Cape, Bellville, Cape Town, South Africa\\
$^{5}$ National Centre for Radio Astrophysics, Tata Institute of Fundamental Research, Pune 411007, India\\
}
\pubyear{2019}
\begin{document}
\label{firstpage}
\pagerange{\pageref{firstpage}--\pageref{lastpage}}
\maketitle
\begin{abstract}
The large-scale CMB B-mode polarization is the direct probe to the low frequency primordial gravitational wave signal. However, unambiguous measurement of this signal requires a precise understanding of the possible contamination. One such potential contamination arises from the patchiness in the spatial distribution of free electrons during the epoch of reionization. We estimate the B-mode power spectrum due to patchy reionization using a combination of \emph{photon-conserving} semi-numerical simulation and analytical calculation, and compare its  amplitude with the primordial B-mode signal.  For a reionization history which is in agreement with several latest observations, we find that a stronger secondary B-mode polarization signal is produced when the reionization is driven by the sources in massive halos and  its amplitude can be comparable to the recombination bump for tensor to scalar ratio $(r) \lesssim 5 \times 10^{-4}$. If contamination from patchy reionization is neglected in the analysis of B-mode polarization data, then for the models of reionization considered in this analysis, we find a maximum bias of about $30\%$ in the value of  $r=\,10^{-3}$ when spatial modes between $\ell \in [50, 200]$ are used with a delensing efficiency of $50\%$. The inferred bias from patchy reionization is not a severe issue for the upcoming ground-based CMB experiment Simons Observatory, but can be a potential source of confusion for proposed CMB experiments which target to detect the value of $r< 10^{-3}$. However, this obstacle can be removed by utilizing the difference in the shape of the power spectrum from the primordial signal.
\end{abstract}
\section{Introduction}
The reionization of cosmic hydrogen is believed to have begun at a redshift $z \sim 15-20$ when the first stars formed. Due to the very large photoionization cross section at  {energy} 13.6 eV, the ultraviolet photons are absorbed by gas in the immediate vicinity of the sources, forming ``bubbles'' of ionized hydrogen. This leads to large spatial fluctuations in the ionized fraction, and reionization is said to be ``patchy''. The bubbles grow and eventually merge, resulting in a uniformly reionized universe at $z \lesssim 6$ \citep{BarkanaLoeb:2001, BarkanaLoeb:2004, FZH:2004, Wyithe:2003, 2018PhR...780....1D}. Intriguing details of the Epoch of Reionization (EoR) have been gleaned by a host of cosmological observables in recent past. 
In particular, the `reionization bump' in the Cosmic Microwave Background (CMB) anisotropy measurements of small \textit{l} polarization constraints the mean value of reionization optical depth $\tau$ which results from the Thompson scattering of CMB photons through interaction with free electrons emanated from reionization. The latest result by the Planck satellite \citep{Planck:2018} reports a total optical depth to recombination to be $\tau = 0.054\pm 0.007$ \footnote{This value changes by about $1$-$\sigma$ with change in the reionization model \citep{Planck:2018}.}. However, due to the patchiness in the process of reionization, secondary anisotropies gets generated in the CMB temperature and polarization at all angular scales \citep{Hu:2000,Santos:2003,Zahn:2005,MCQUINN:2006,Mortonson:2007, Dore:2007, Dvorkin:2008tf,Dvorkin:2009,2011arXiv1106.4313S,2013PhRvD..87d7303G}.

Future CMB missions aim to make exquisite measurement of secondary CMB anisotropies due to various physical effects such as gravitational lensing \citep{Seljak:1995ve, Lewis:2006fu,Hu:2001, Hu:2002}, integrated Sachs-Wolfe \citep{Sachs:1967er},  thermal Sunyaev~-Zel’dovich (tSZ) effect \citep{SZ:1970}
, kinetic Sunyaev~-Zel’dovich effect (kSZ) \citep{Sunyaev:1980nv, SZ:1980, Nozawa:1998} 
and patchy reionization \citep{Dvorkin:2008tf,Dvorkin:2009,2011arXiv1106.4313S,Battaglia:2012id,Battaglia:2012im, 2013ApJ...776...82N, 2013PhRvD..87d7303G, Park:2013mv, Alvarez:2015xzu, Namikawa:2017uke,Roy:2018}. A possible source of secondary anisotropy is the B-mode polarization arising from the patchy electron distribution during reionization. This can potentially contaminate the recombination bump of the B-mode polarization leading to a bias in the inferred value of $r$. Hence an accurate and reliable computation of the signal arising from reionization is crucial for the upcoming experiments that aim to detect the primordial B-modes \citep{Matsumura:2013aja, Ade:2018sbj, 2016arXiv161002743A, 2018SPIE10698E..46Y}. As per the current understanding, reionization is an inhomogeneous and complex process \citep{Mesinger:2007, Zahn:2007, Choudhury:2009, Mitra:2016olz} which depends on various astrophysical processes related to early galaxy formation.

This work aims to study the contamination from patchy reionization on the inferred value of tensor to scalar ratio ($r$) and tensor spectral index $(n_t)$. We estimate the signal of patchy reionization by calculating the inhomogeneous electron density using a combination of photon-conserving semi-numerical simulations \citep{Choudhury:2018} and analytical calculations. 
By using the power spectrum of the electron density from the semi-numerical and analytical method, we estimate the CMB B-mode polarization power spectrum due to patchy reionization for different reionization scenarios. We then estimate the implication of the patchy reionization on the value of $r$, which are accessible from the upcoming and proposed CMB experiments such as LiteBIRD \citep{Matsumura:2013aja}, Simons Observatory (SO) \citep{Ade:2018sbj}, CMB Stage-4 (CMB-S4) \citep{2016arXiv161002743A}, CMB-Bharat  \footnote{Space-based CMB mission proposed by India to Indian Space Research Organization (ISRO) (\url{http://cmb-bharat.in/}).} and Probe of Inflation and Cosmic Origins (PICO) \citep{2018SPIE10698E..46Y,Hanany:2019lle}. Our analysis indicates that in order to make an unambiguous  {detection of $r\leq10^{-3}$}, we need to remove the patchy reionization signal (de-tau) along with the lensing (de-lensing) for proposed CMB experiments such as CMB-S4, CMB-Bharat and PICO.

In section \ref{formalism}, we discuss the effect of patchy reionization on CMB B-modes. In section \ref{elec_den}, we estimate the electron density power spectrum using analytical and \textit{photon conserving} semi-numerical method. The estimation of the B-mode polarization signal due to patchy reionization and its effect on $r$ and $n_t$ are discussed in section \ref{signal_est} and section \ref{rcont} respectively. Finally we conclude the main findings of this analysis and future  {directions} in section \ref{conclusion}. Throughout this work, we use the usual flat $\Lambda$CDM cosmology with parameters given by $\Omega_m = 0.308$, $\Omega_b = 0.0482$, $h = 0.678$, $n_s = 0.961$, and $\sigma_8 = 0.829$ \citep{2014A&A...571A..16P} which are consistent with \citep{Planck:2018}.

\section{Formalism of  Large angular scale CMB B-mode polarization from patchy reionization}\label{formalism}
CMB polarization can be expressed by the Stokes parameters $Q$ and $U$, which behaves like a spin-2 field on the sphere and under rotation by angle $\phi$ transforms as \citep{Zaldarriaga:1996xe,Kamionkowski:1996ks}
\begin{equation}\label{pol-def}
(Q \pm iU)(\hat n')= (Q \pm iU)(\hat n')e^{\mp2 i \phi}.
\end{equation} 
 {They can be decomposed} into the rotation invariant $E$-mode and B-mode polarization field as \citep{Zaldarriaga:1996xe,Kamionkowski:1996ks}
\begin{eqnarray}\label{eb-def}
E(\hat n)&=& \frac{-1}{2} \bigg[\partial^2_L (Q + iU)(\hat n) + \partial^2_R (Q - iU) (\hat n)\bigg], \nonumber \\
&=&\sum_{lm} \bigg[\frac{(l+2)!}{(l-2)!}\bigg]^{1/2}a^E_{lm} Y_{lm}(\hat n),\nonumber\\
B(\hat n)&=& \frac{i}{2} \bigg[\partial^2_L (Q + iU)(\hat n) - \partial^2_R (Q - iU) (\hat n)\bigg], \nonumber \\
&=&\sum_{lm} \bigg[\frac{(l+2)!}{(l-2)!}\bigg]^{1/2}a^B_{lm} Y_{lm}(\hat n),
\end{eqnarray}
where $\partial_R$ and $\partial_L$ are the spin raising and lowering differential operators respectively \citep{Zaldarriaga:1996xe,Kamionkowski:1996ks}. Using Eq. \ref{eb-def}, we can write the $E$ and $B$ polarization field in the spherical harmonic basis as
\begin{eqnarray}\label{e-filed}
a^E_{lm} &=& \frac{-1}{2}\left[{}_{+2}a^{pol}_{lm} + {}_{-2}a^{pol}_{lm}\right], \nonumber \\
a^B_{lm} &=& \frac{i}{2}\left[{}_{+2}a^{pol}_{lm} - {}_{-2}a^{pol}_{lm}\right],
\end{eqnarray}
where ${}_{\pm2}a^{pol}_{lm}= \int d^2\hat n {}_{\pm2}Y^*_{lm}(\hat n) (Q \pm iU)(\hat n)$.

The  {primordial} $E$-mode gets generated by both scalar and tensor perturbations, whereas the  {primordial} B-mode gets generated only by tensor perturbations. The secondary anisotropies give rise to both $E$-mode and B-mode polarization  {and} can act as a potential source of contamination to measure the primary CMB polarization signal.
A possible source of secondary anisotropy to the polarization field arises during the epoch of reionization due to spatial inhomogeneities in the distribution of free electrons \citep{Hu:2000,Santos:2003,Zahn:2005,MCQUINN:2006,Mortonson:2007, Dvorkin:2008tf,Dvorkin:2009,2011arXiv1106.4313S}. Spatial inhomogeneities during the epoch of reionization, can generate secondary polarization due to two processes (i) \textit{Scattering:} Thompson scattering of the CMB quadrupole by the free electrons during the epoch reionization \citep{Hu:2000,Liu:2001xe,Mortonson:2007, Dore:2007,  Dvorkin:2008tf}, (ii) \textit{Screening:} The primary E-mode polarization gets converted into B-mode polarization due to anisotropic optical depth  \citep{Dvorkin:2009}.
The first effect is important at large angular scales, while the screening effect becomes dominant at small angular scales \citep{Dvorkin:2009}. In this paper, we focus on the contamination to the primordial B-modes at large angular scales due to the effect from \textit{scattering}.

The angular variation in the free electron distribution at a comoving distance $\chi$  can be written as a sum of the global mean and the fluctuating component
\begin{eqnarray}\label{xe-def}
x_e (\hat n, \chi)= \bar x_e(\chi) + \Delta x_e(\hat n, \chi),
\end{eqnarray}
where $x_e \equiv n_e / n_H$ is the free electron fraction with respect to the hydrogen. This leads to a direction dependence in the photon optical depth $\tau(\hat n, \chi)$, which can be written as
\begin{eqnarray}\label{tau-def}
\tau(\hat n, \chi)= \sigma_T \bar{n}_{H} \int_0^{\chi} d \chi' \left(1 + z'\right)^{ {2}} x_e(\hat n, \chi'),
\end{eqnarray}
where $\bar{n}_H$ is the mean comoving density of hydrogen and $\sigma_T$ is the Thomson scattering cross-section.
Photons passing through the patchy electron density field pick up additional polarization fluctuations which can be written as \citep{Hu:1997hp,Hu:2000}
\begin{eqnarray}\label{pol-anis}
(Q \pm i U)(\hat n)= \int d\chi ~\dot\tau(\hat n, \chi)e^{-\tau(\hat n, \chi)} \mathcal{S}^\pm(\hat n, \chi),
\end{eqnarray}
where $S_{pol}(\hat n, \chi)$ is the source term for the polarization, which is related to the local temperature quadrupole as
\begin{eqnarray}\label{cmb-quad}
\mathcal{S}^\pm(\hat n, \chi)= \frac{-\sqrt{6}}{10}\sum_m ({}_{\pm2}Y_{2m}(\hat n)) ~a^T_{2m} (\hat n, \chi).
\end{eqnarray}

The corresponding B-mode polarization power spectrum from the scattering effect can be written as \footnote{The quadrupole anisotropy of temperature $(C^{T}_2)^{1/2}= \sqrt{\frac{4\pi}{5}}Q_{RMS}$ is used in this analysis \citep{Hu:2000}.}
\begin{eqnarray}\label{bb-power-exact}
C_l^{BB}&=& \frac{24\pi \bar{n}^2_{H}\sigma^2_T}{100} \int d\chi \frac{1}{a^2}\int d\chi' \frac{1}{a'^2}e^{-\tau(\chi)- \tau(\chi')} \nonumber \\
&& \int dk \frac{k^2}{2\pi^2} P_{ee} (k, \chi, \chi') j_l(k\chi)j_{l}(k\chi') \frac{Q_{RMS}^2}{ {2}},
\end{eqnarray}
where, we have written the power spectrum of the spatial fluctuation of the electron density as $\langle \Delta x_e (k,\chi')  \Delta x_e^* (k',\chi')\rangle  \equiv P_{ee} (k, \chi, \chi') \delta(k-k')$, $ {Q_{RMS}^2}$ is the  {local} quadrupole temperature  {variance, which we assumed to be constant at a value of $22 \, \mu$K over the redshift range of reionization \citep{Dvorkin:2008tf}} and $j_l(k\chi)$ are the spherical Bessel functions. The above expression of the B-mode power spectrum is valid at all angular scales   {under the approximation} of constant source (or slowly varying source \cite{Hu:2000}).  {Apart from this scattering effect, the screening effect (which is not important at these angular scales \citep{Dvorkin:2009ah}) arises when the direction dependence of  $\tau$ is included in the exponential.}

In order to speed up the numerical calculation, we  incorporate Limber approximation \citep{1953ApJ...117..134L, LoVerde:2008re} to the above expression resulting into the following form of the B-mode power spectrum, valid at large angular scales $l\gtrsim30$  
 \begin{eqnarray}\label{bb-power-limber}
C_l^{BB}= \frac{6 \bar{n}^2_{H}\sigma^2_t}{100} \int \frac{d\chi e^{-2\tau(\chi)}}{a^4\chi^2}
P_{ee} \left(k=\frac{l+1/2}{\chi}, \chi\right)\frac{Q_{RMS}^2}{ {2}}.
\end{eqnarray}
The above expressions  {given in Eq. \ref{bb-power-exact} and \ref{bb-power-limber}} indicate that the power spectrum of the B-mode signal depends on the power spectrum of the electron distribution, which is not a well known quantity due to several astrophysical uncertainties during reionization.

\section{The electron density power spectrum}\label{elec_den}
\begin{figure}
\centering
\includegraphics[trim={0cm 0cm 0cm 0cm}, clip, width=1.\linewidth]{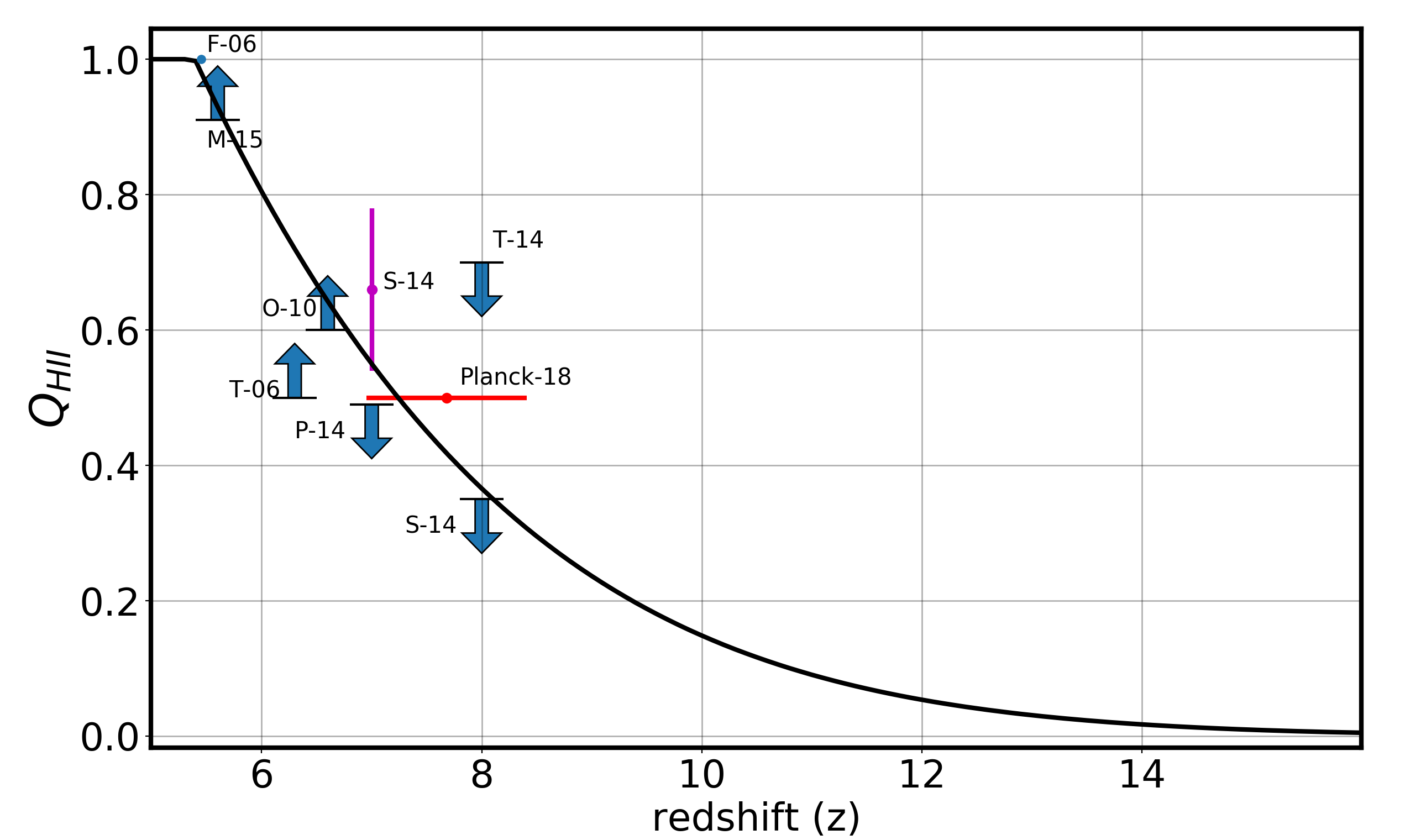}
\caption{Reionization history used in this analysis is depicted along with multiple observations as measured by several probes (\citep{2006AJ....132..117F} (F-06), \citep{2015MNRAS.447..499M} (M-15), \citep{2010ApJ...723..869O} (O-10), \citep{Pentericci:2014nia} (P-14), \citep{Planck:2018} (Planck-18), \citep{Schenker:2014tda} (S-14),  \citep{2014ApJ...794....5T} (T-14), \citep{Totani:2005ng} (T-06)).}
\label{iofrac}
\end{figure}

In this section, we describe the theoretical model used to estimate the power spectrum $P_{\rm ee}(k)$ of  {$\Delta x_e$}. 
The fluctuations in the electron density is driven by the ionized bubbles produced during the epoch of hydrogen reionization. We can write the free electron density at a 
 {redshift $z$} and spatial position $\textbf{x}$ in terms of the comoving density $n_{\rm HII}$ of ionized hydrogen (HII) as
\begin{eqnarray} \label{nevalue}
n_e(\textbf{x}, z) &=& \chi_{\rm He}~n_{\rm HII}(\textbf{x}, z) \nonumber \\
&=& \chi_{\rm He}~x_{\rm HII}(\textbf{x}, z)~n_H(\textbf{x}, z) \nonumber \\
&=& \chi_{\rm He}~\bar{n}_H~x_{\rm HII}(\textbf{x}, z)~\Delta_H(\textbf{x}, z),
\end{eqnarray}
where $\chi_{\rm He} = 1.08$ accounts for excess free electrons because of the presence of singly-ionized helium, $x_{\rm HII}$ is the ionized fraction of hydrogen, 
and $\Delta_H$ is the overdensity of hydrogen. At large scales, we assume the hydrogen density fluctuations to trace the underlying dark matter field $\Delta_H = 1 + \delta_{\rm dm}$, where $\delta_{\rm dm}$ is the dark matter density contrast. The globally averaged value of the electron density is $\bar{n}_e(z) = \chi_{\rm He}~\bar{n}_H~Q_{\rm HII}(z)$, where $Q_{\rm HII}(z) \equiv \langle x_{\rm HII}(\textbf{x}, z)~\Delta_H(\textbf{x}, z) \rangle$ is the mass-averaged ionized fraction. The electron fraction is thus given by
\begin{equation}
x_e(\textbf{x}, z) = \frac{n_e(\textbf{x}, z)}{\bar{n}_H} = \chi_{\rm He}~x_{\rm HII}(\textbf{x}, z)~\Delta_H(\textbf{x}, z),
\end{equation}
which is used for calculating the power spectrum $P_{ee}(k, z)$.

In this work, we simulate the large-scale dark matter density field using the publicly available $N$-body code GADGET-2 \citep{GADGET2:2005}, with initial conditions generated using N-GENIC\footnote{\tt \url{https://wwwmpa.mpa-garching.mpg.de/gadget/n-genic.tar.gz}}. The simulation was performed on a large 512$h^{-1}$Mpc size cubical box containing $256^3$ collisionless dark matter particles.

Since the simulation is not of high resolution, we cannot use the direct group-finder algorithms to identify the collapsed haloes responsible for producing the ionizing photons. We rather employ a method based on the conditional mass function obtained from the ellipsoidal collapse as outlined in \citep{Seehars:2015ada, Choudhury:2018}. The net result of this exercise is that, for every grid cell in the simulation box, we have the values of the density fluctuations $\Delta_H$ and the fraction $f_{\rm coll}(>M_{\rm min})$ of mass in collapsed objects of mass $> M_{\rm min}$.

Given the density and collapse fraction fields along with the value of the ionizing efficiency $\zeta$ (which, in principle, can be a function of $z$), one can generate the ionization field $x_{\rm HII}$ efficiently using semi-numerical techniques. The conventional excursion-set models have been extremely useful in such cases \citep{Mesinger:2007, 2011MNRAS.411..955M, Zahn:2007,Santos:2007dn,Geil:2007rj,Choudhury:2009}. However, they have been known to have two crucial shortcomings, namely, (i) they violate photon-conservation in the sense that the number of hydrogen atoms ionized (accounting for recombinations) is not equal to the number of photons produced by the sources and (ii) more importantly, the power spectrum of the ionized field at large scales does not converge to  {the} same value for different grid resolutions. The second shortcoming can have severe implications for using these models while comparing with observations.

Hence in this work we rather employ a different algorithm based on \citep{Choudhury:2018}, where the ionization field is generated using an explicitly photon-conserving algorithm. This algorithm is found to be reasonably efficient and produces a large-scale ionization field that is independent of the resolution. 

We fix the reionization history for different values of $M_{\rm min}$ as described below.  {We will use this parameter to label different reionization models in this analysis.}

\begin{itemize}

\item As the first step, we choose $M_{\rm min} = 10^8 M_\odot$, appropriate for atomically cooled gas in haloes, and assume $\zeta$ to be $z$-independent. We fix the value of $\zeta$ in a  way that the resulting reionization history is consistent with the $\tau$-constraints given by \citep{Planck:2018}. The resulting ionization history, which gives $\tau = 0.055$, is shown in Figure \ref{iofrac}. In the figure, we also show various direct and indirect measurements of $Q_{\rm HII}$. It is clear from the figure that our reionization history is consistent with all available constraints. This evolution of $Q_{\rm HII}$ is also consistent with the calculations of \citep{Kulkarni:2018erh} which explain the large-scale fluctuations observed in the quasar absorption spectra at $z \sim 5.5$.

\item We also consider higher values of $M_{\rm min}$, namely $10^9, 10^{10}$ and $10^{11} M_\odot$. The case $M_{\rm min} = 10^9 M_\odot$ corresponds to the case where star-formation in low-mass haloes is suppressed by radiative feedback. The extreme case of $M_{\rm min} = 10^{11} M_\odot$ corresponds to ionizing photons sourced by extremely rare haloes with circular velocities $\sim 250$~km~s$^{-1}$. Such a scenario will be consistent with reionization driven by AGNs \citep{Kulkarni:2017qwu}. Though such models are in tension with various low-redshift data \citep{Mitra:2016olz}, we nevertheless consider them in this work as extreme limiting cases.

\item For each value of $M_{\rm min}$, we choose $\zeta$ at each redshift so that $Q_{\rm HII}$ is consistent with the reionization history in Figure \ref{iofrac}. Since haloes of larger mass are relatively rarer at higher redshifts, we require a $\zeta$ that decreases with decreasing redshift.

\end{itemize}

Using the above algorithm, we can calculate $P_{ee}(k, z)$ from the simulation box for $k \gtrsim 0.01~h$~Mpc$^{-1}$. For our redshifts of interest, this allows us to estimate $C_l^{BB}$ at angular multipoles $l \sim k \chi \gtrsim 80$. 

For studying the signal  {at large} angular scales, we thus need to use some other method. It can be shown that the ionization fluctuations at scales larger than the bubble sizes are determined by the fluctuations in the density field of the haloes which produce the ionizing photons. In fact, it is straightforward to show that the power spectrum of electron density field at large scales is 
\begin{eqnarray}\label{xepower}
P_{ee}(k, z, z')= \mathcal{R}( {k},z) \mathcal{R}( {k},z')D(z)D(z')P_{L, DM}(k),
\end{eqnarray}
where we have defined $\mathcal{R} ( {k}, z)\equiv \chi_{\rm He} Q_{\rm HII} (z)b_{h}(k, M_{\rm min}, z)$ for $Q_{\rm HII} < 1$, with $b_h(k, M_{\rm min}, z)$ being the mass-weighted halo bias for haloes more massive than $M_{\rm min}$. At large scales, the dark matter power spectrum is given by the linear theory with $D(z)$ being the linear growth factor.  {This equation remains valid at scales larger than the largest bubbles and hence this approximation will violate towards the end of reionization. However, towards the end of reionization, electron density will become homogeneous and the amount of patchy reionization signal arising will also be less. So, we do not expect to miss any significant signal of patchy reionization by making this assumption.}   {We compute the value of $\mathcal{R}( {k},z)$ from the simulation box at the largest available scales}. Note that this prescription works well only when the scales probed are larger than the bubble sizes, hence the relation becomes less and less accurate towards late stages of reionization when the bubbles become comparable to the simulation box. At the same stage of reionization, the above relation is least accurate for large $M_{\rm min}$, and hence our results are less robust at large scales for the $M_{\rm min} = 10^{11} M_\odot$ case.

\begin{figure}
\centering
\includegraphics[trim={0.3cm 0.3cm 0.3cm 0.3cm}, keepaspectratio=True, clip,width=0.5\textwidth]{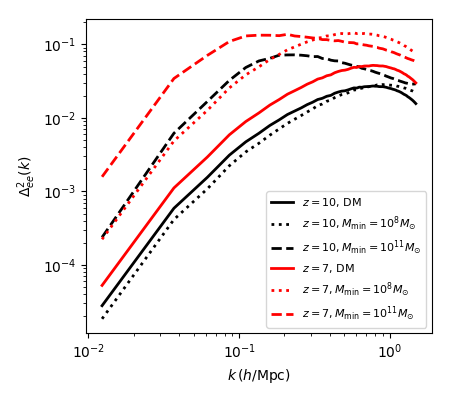}
\caption{ {We plot the dimensionless electron power spectrum $\Delta_{ee}^2 (k)\equiv k^3\rm{P}_{\Delta x_e \Delta x_e}/(2\pi^2)$ for two different redshifts ($z=7$ and $z=10$) and minimum halo masses ($10^{8}\, M_\odot$ and $10^{11}\, M_\odot$) from the photon-conserving simulation of reionization which are used in this analysis. For comparison, we also plot the dimensionless dark matter linear power spectrum for the same values of the redshift.}}
\label{Pmap}
\end{figure}

We show the dimensionless electron power spectrum $\Delta^2_{ee}(k) \equiv k^3 P_{ee}(k) / 2 \pi^2$ in Figure \ref{Pmap}. The power spectra clearly show a peak at scales corresponding to the typical bubble sizes. It is clear from the figure that the bubble sizes grow as reionization progresses (i.e., with decreasing redshift), and are also larger for larger $M_{\rm min}$.

\section{Estimation of the B-mode power spectrum  {from patchy reionization}}\label{signal_est}
\begin{figure*}
\centering
\includegraphics[trim={0.cm 0.cm 1cm 0cm}, keepaspectratio=True, clip,width=.85\textwidth]{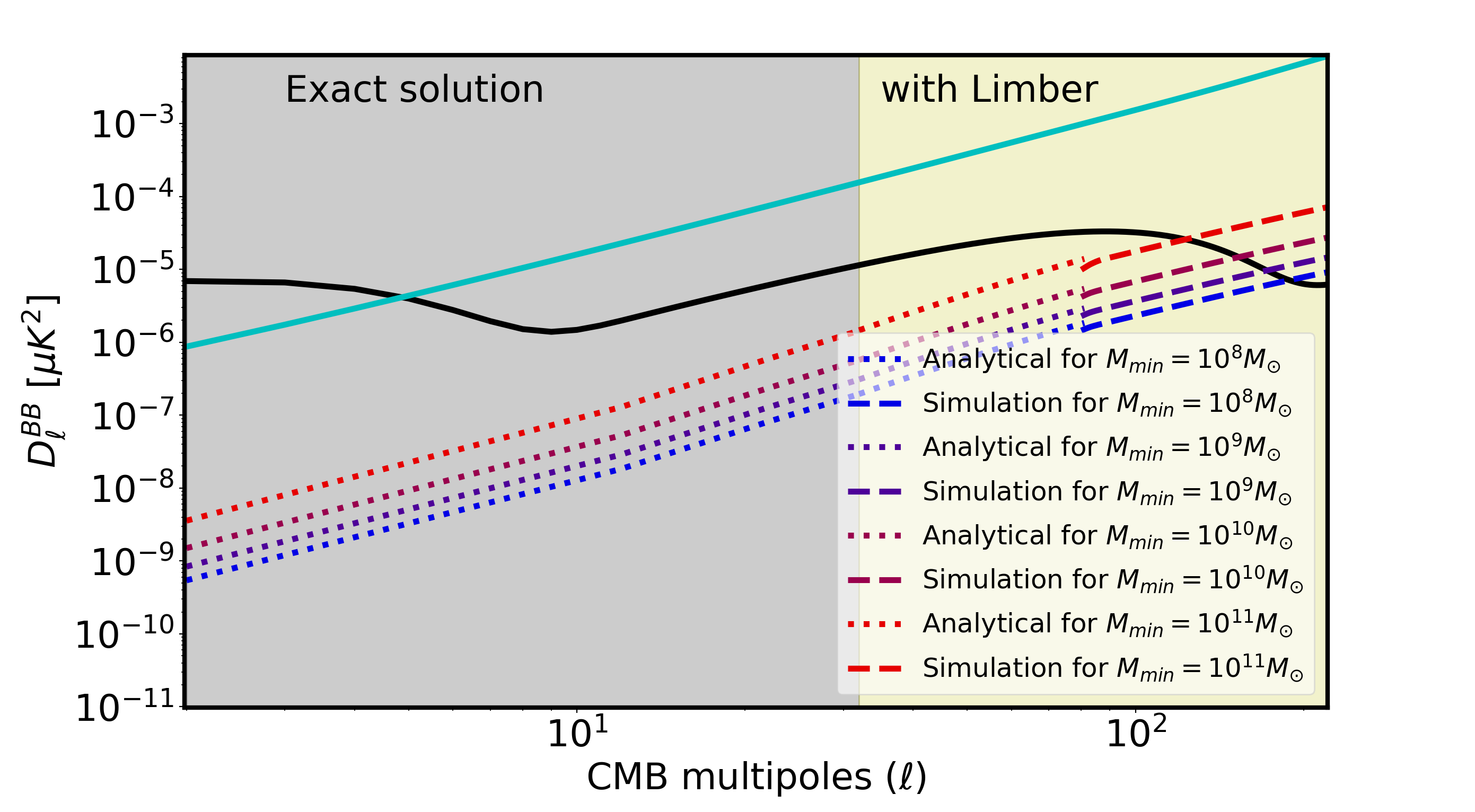}
\caption{The angular power spectrum of the B-mode polarization signal ($D_l^{BB}\equiv l(l+1)C_l^{BB}/2\pi$) are shown for different minimum mass of the halos. The grey and yellow shaded regions denote the  {exact} results ({without Limber approximation}) and the results with Limber approximation respectively.  {The cyan and black solid lines denote the B-mode polarization power spectrum due to weak lensing (with the lensing amplitude $A_{lens}=0.5$) and primordial gravitational waves with tensor to  scalar ratio $r=5 \times 10^{-4}$ respectively.}}
\label{C_lBBplot}
\end{figure*} 
Using the estimate of the electron density power spectrum given in section \ref{elec_den}, we depict the estimated B-mode power spectrum from patchy reionization in Figure \ref{C_lBBplot}. By using the analytical estimate of the electron density power spectrum, we obtain the power spectrum of the B-mode polarization signal at large angular scales ($l< 80$) which are shown by the  {dotted}-lines. For $l<32$, the analytical estimates are obtained by using the integral given in Eq. \ref{bb-power-exact}, whereas for $l>32$, we used  Limber approximation to estimate the B-mode signal (given in Eq. \ref{bb-power-limber}). For $l \gtrsim 80$, we estimate the signal from simulations with Limber approximation and  {the results are} shown by the dashed-lines. For comparison, we also plot the primordial gravitational waves signal for tensor to scalar ratio $r= 5 \times 10^{-4}$  {and the B-mode power spectrum due to weak lensing with $50\%$ delensing ($A_{lens}=0.5$) in black and cyan solid line respectively} using CAMB \citep{Lewis:1999bs, Howlett:2012mh}.

Our estimates show that at the angular scales corresponding to  {the} recombination bump, the secondary B-mode polarization from patchy reionization can be  {another source of contamination to the primordial gravitational waves signal ($r \sim 5 \times 10^{-4}$) along with other contamination such as weak lensing and astrophysical foregrounds.} The amplitude of the signal depends strongly on the mass of the halos $M_{min}$ which drive reionization. The amplitude from patchy reionization can vary with $M_{min}$ by order of magnitude even when the ionization fraction is fixed (shown in Figure \ref{iofrac}) and obeys the recent observational constraints \citep{Bouwens:2015vha, Planck:2018}.  {The signal from patchy reionization is stronger when the reionization is driven by massive halos which create bigger bubbles during the epoch of reionization.  {Our conclusion also agrees with the previous estimates of the patchy-reionization B-mode polarization signal by \citep{Mortonson:2007}. Using an analytical model (log-normal distribution) of ionized bubble radius, it was shown that the reionization driven by bigger bubbles produces stronger B-mode signal than for the case with smaller bubbles.}

\section{Contamination to tensor to scalar ratio and tensor spectral index due to patchy reionization}\label{rcont}
\begin{figure*}
\centering
\subfloat[]{\includegraphics[trim={0.5cm 0.cm 1cm 0cm}, keepaspectratio=True, clip,width=0.5\textwidth, 
]{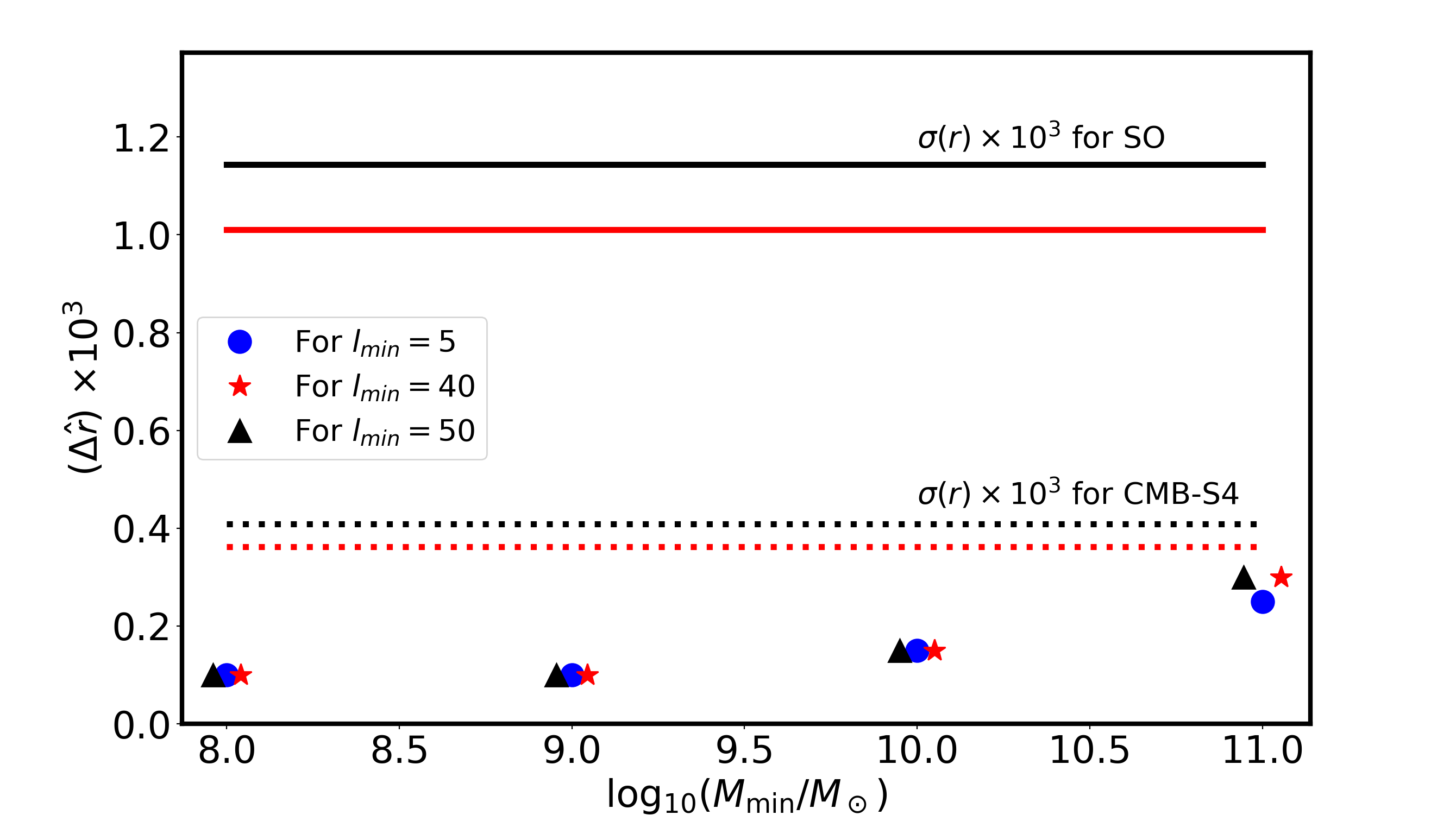}\label{rbias}}
\subfloat[]{\includegraphics[trim={0.5cm 0.cm 1cm 0cm}, keepaspectratio=True, 
clip,width=0.55\textwidth]{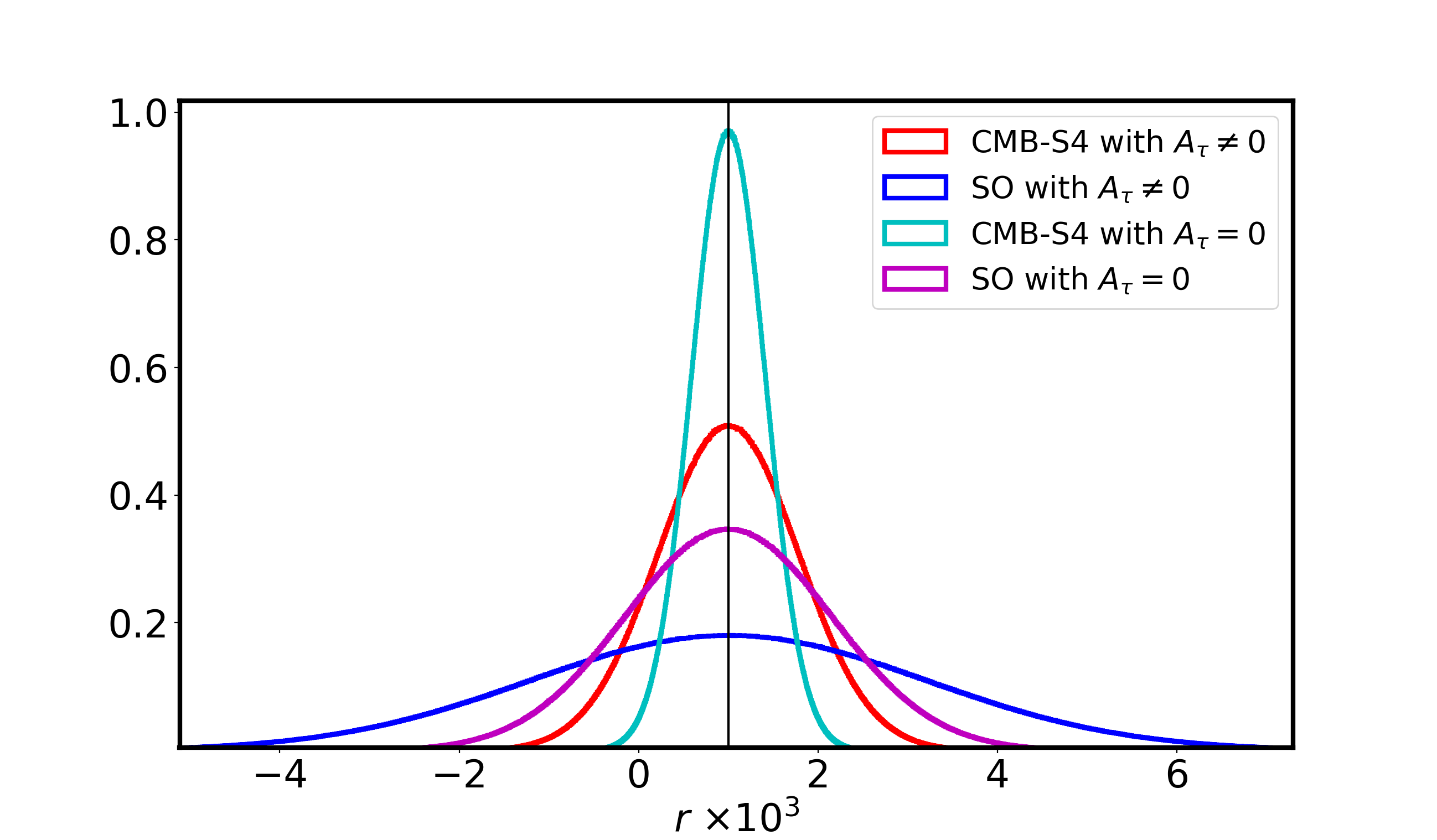}\label{r1d}}
\caption{ {Effects of patchy reionization on the mean and variance of r with $50\%$ delensing. Assuming $r_{true} = 10^{-3}$. (a) Bias in the inferred Maximum Likelihood value of $r$ for different scenarios of patchy reionization and $l_{min}$ cutoff for Simons Observatory (SO)
and CMB-S4. The lines represent $1$--$\sigma$ error-bars for SO (solid-lines) and CMB-S4 (dotted) for $l_{min} = 40$ (red) and $l_{min} = 50$ (black). For visual purposes, the triangle and
star shaped markers are shifted in the x-coordinate. (b) The posterior probability distribution after marginalizing over $A_\tau$, centered on the assumed fiducial value. For both SO and CMB-S4, assuming $M_{min} = 10^{10} M_\odot$. For comparison, the distributions in the absence of patchy reionization are also shown.}}
\label{rbiasall}
\end{figure*} 

The B-mode signal from patchy reionization will lead to a bias in the mean value of the  {inferred} tensor to scalar ratio ($r$) and the spectral index of the tensor perturbations ($n_t$). In order to quantify the bias in the values of $r$ and $n_t$, we make a Maximum Likelihood (ML) estimation of $r$ and $n_t$ using the  {form of the  log-likelihood} ($\mathcal{L}$)
\begin{equation}\label{likelihood}
-2\mathcal{L} \propto \sum_{l, l'=l_{min}}^{l_{max}} (\tilde C^{BB}_l - C^{BB}_l)\Sigma^{-1}_{ll'}(\tilde C^{BB}_{l'} - C^{BB}_{l'}),
\end{equation}
 {where, $\tilde C^{BB}_l$ and $C^{BB}_l$ are the mock data power spectrum and model power spectrum respectively.} The mock data power spectrum can be written as
\begin{equation}
\tilde C^{BB}_l= C^{BB,prim}_l + A_{\tau}C^{BB, reion}_l (M_{min}) + A_{lens}C^{BB, lens}_l,
\end{equation}
which includes primordial CMB $C^{BB,prim}_l$ (at the fiducial value $r=10^{-3}$ and $n_t=0$), patchy reionization signal for minimum halo mass denoted by $C^{BB, reion}_l (M_{min})$ and B-mode signal from weak lensing $C^{BB, lens}_l$.  $A_{lens}$ indicates the residual lensing amplitude after  delensing the lensing signal. In our analysis, we have taken $A_{lens}= 0.5$ which is possible to achieve using several delensing algorithms   \citep{Simard:2014aqa,Sherwin:2015baa, Carron:2017vfg,PhysRevD.96.123511, Manzotti:2017net, Millea:2017fyd}.   $A_{\tau}$ denotes the normalization chosen to include or exclude the patchy reionization signal. 

In Eq. \ref{likelihood}, $C^{BB}_l$ denotes the model  {(or the theoretical)}  {power spectrum} of B-mode signal estimated using CAMB \citep{Lewis:1999bs, Howlett:2012mh}, with an assumed delensing efficiency  $A_{lens}=0.5$. In order to show the effect of patchy reionization on the inferred value of $r$ and $n_t$, we have not considered patchy reionization in the model power spectrum  $C^{BB}_l$. $\Sigma_{ll'}$ is the covariance matrix of the B-mode power spectrum which can be written as 
\begin{equation}
\Sigma_{ll'}= \frac{2}{f_{sky}(2l+1)}\left(\tilde C^{BB}_l + N_l\right)^2\delta_{ll'},
\end{equation}
where $N_l$ is the power spectrum of the instrument noise. In this analysis we have used the instrument noise of Simons Observatory (SO) \citep{Ade:2018sbj} and CMB-S4 \citep{2016arXiv161002743A} \footnote{The exact instrument noise for CMB-S4 are yet to be finalized.} for the frequency channel $90$ GHz and $150$ GHz respectively.  For SO, we have used the goal polarization noise \footnote{$\sqrt{2}$ times the noise in intensity.} (for the Small Aperture Telescopes (SATs) with sky-fraction $f_{sky}=0.1$) as $2.7 \mu K_{CMB}-$arcmin with Full Width Half Maximum (FWHM) of the beam as $30$-arcmin. For CMB-S4 we have used polarization noise  as $2.5 \mu K_{CMB}-$arcmin with FWHM of the beam as $1.6$-arcmin and sky-fraction $f_{sky}=0.7$. We have ignored the additional noise due to foreground contamination in this analysis and have assumed that the foreground contamination can be cleaned using the remaining five frequency channels of SO and four frequency channels of CMB-S4. The ground-based experiments will not have access to the large angular scales or the low multipoles ($l$) values, unlike the space-based missions. As a result, we have included an additional parameter $l_{min}$ which sets the minimum value of CMB multipole which can be considered in the measurement. We choose CMB multipoles up to $l_{max}=200$ for three different $l_{min}$ cutoffs to obtain the bias in the values of $r$ and $n_t$.   

\textit{\textbf{Effect on $\mathbf{r}$}} : The  difference between the ML estimated value  {for $r$ when neglecting $A_\tau$} in the analysis and the fiducial injected value of $r=10^{-3}$ is shown in Figure \ref{rbias} for different choices of $l_{min}$ and minimum halo mass $M_{min}$ parameters. 
The bias $\Delta r$ in the measurement of $r=10^{-3}$ is about  {$30 \%$} for $M_{min}= 10^{11}\, M_\odot$ and decreases as we go to smaller values of  {$M_{min}$}.  {Our results are robust to the transition (discontinuity) from the analytical to simulation results. We have validated the robustness of our results by using only the analytical model for the complete $l$ range. For the values of $l\leq 100$ we find perfect agreement between the results obtained from analytical and analytical-simulation hybrid model. For the values of $l$ in the range $[100, 200]$, analytical model mildly overestimates the fluctuations than the simulation, resulting into $\sim 10\%$ increment in the bias $\Delta \hat r$ in comparison to the simulation case.}
The bias gets stronger for the higher value of $l_{min}$. This happens because the contribution from patchy reionization affects more strongly the recombination bump than the reionization bump. As a result, experiments which cannot measure reionization bump at low $l$ and relies only on the recombination bump to measure the value of $r$, gets more affected by patchy reionization. So, space-based CMB missions are less susceptible to the effects from patchy reionization than the ground-based CMB experiments.  {For different values of $l_{min}$, we explore the range of $l$ which drives the contribution in $\Delta \hat r$. For both ground-based and space-based CMB experiments, about $85\%$ contribution of its $\Delta \hat r$ comes from the $l$ range $[50, 90]$  and $[5, 90]$ respectively. The remaining $15\%$ of $\Delta \hat r$ arises from the values of $l>90$. }

So, in order to make a robust measurement of the value of  $r$, we need to remove the patchy reionization signal (de-tau) from the primordial gravitational waves and develop an optimal estimator which can infer the signal of both primordial gravitational waves and patchy reionization jointly from the CMB data. This is possible to construct as the shape of the B-mode power spectrum is different for patchy reionization and primordial gravitational waves (as shown in Figure \ref{C_lBBplot}). The shape of the B-mode power spectrum from primordial gravitational waves is expected to be robust within the standard $\Lambda$CDM model of cosmology. But the B-mode power spectrum from patchy reionization is susceptible to vary due to several astrophysical uncertainties. 
From the semi-numerical and analytical methods considered in this analysis, we find that the shape of the B-mode power spectrum for different reionization scenarios mainly differs in the amplitude, but the shape of the power spectrum remains roughly constant over the range $l \sim 2-200$. We expect this to hold if the typical ionized bubbles are smaller than the angular scales (corresponding to these $l$ range) and these bubbles follow the linear dark matter distributions. However, this assumption gets violated towards the end of reionization, and analyses from larger simulation box-size are required to capture the shape of the B-mode power spectrum more accurately.
  
Along with the bias in the mean value of $r$, the effect from patchy reionization will also increase  the error-bar  on $r$ denoted by $\sigma(r)$. We obtain a Cramer-Rao bound  $\sigma(r) \geq (F^{-1})_{rr}^{1/2}$ on the value of $r$ after marginalizing the Fisher matrix $\mathbf{F}$ over the patchy reionization amplitude $A_\tau$ at a fiducial value of patchy reionization corresponding to $M_{min}=10^{10} M_\odot$, where the Fisher matrix $\mathbf{F}$ can be written as \citep{Tegmark:1996bz}
\begin{equation}
F_{\alpha \beta}= -\bigg\langle \frac{\partial^2\mathcal{L}}{\partial p_\alpha\partial p_\beta}\bigg\rangle.
\end{equation}
In our estimate, we have taken two parameters ($p_i \in$ ($r$ and $A_{\tau}$) with a fixed value of $n_t=0$ and $l_{min}=50$ and $l_{max}=200$. The comparison of the Gaussian probability distribution function for the cases with ($A_{\tau}=1$) and without ($A_{\tau}=0$) patchy reionization for a fiducial value of $r=10^{-3}$ are shown in Figure \ref{r1d} for SO and CMB-S4.   

\textit{\textbf{Effect on $\mathbf{n_t}$}:} Tensor spectral index $n_t$ gets biased towards a positive value resulting in a blue spectrum when patchy reionization is not considered in the theoretical modelling. The bias in the value of $n_t$ is nearly independent of the value of $M_{min}$ and gives roughly a constant shift in the inferred value $\hat n_t =0.04$ for all the scenarios considered in this analysis. So, even though several inflationary theories predict $n_t=-r/8$, patchy reionization can lead to a blue spectrum of the tensor perturbations. Hence, it is essential to understand the signature of patchy reionization very accurately in order to make an unbiased estimate of $n_t$. However, the error-bar on $n_t$ from upcoming CMB ground-based missions such as SO and CMB-S4 are larger than the typical value of $n_t$ due to patchy reionization (for the models considered in this analysis) and as a result, this effect cannot be measured from SO and CMB-S4.

\section{Conclusion}\label{conclusion}
In this paper, we explore the contamination to primordial gravitational wave signal due to secondary CMB anisotropies sourced by patchy reionization. Our study based on an \emph{explicitly photon-conserving} semi-numerical simulation of the epoch of reionization \citep{Choudhury:2018} indicates that different history of reionization can generate B-mode  {power spectrum with amplitudes varying by an order of magnitude.} If reionization is driven by larger halo masses, then the RMS fluctuations in the patchy reionization signal are larger and hence causes a large B-mode power spectrum.  This in turn can cause a large bias in the inferred value of the tensor to scalar ratio ($r$) if the contamination from the patchy reionization is unknown and not considered in the analysis.

\begin{figure}
\centering
\includegraphics[trim={0.cm 0.cm 0cm 0cm}, keepaspectratio=True, clip,width=0.5\textwidth]{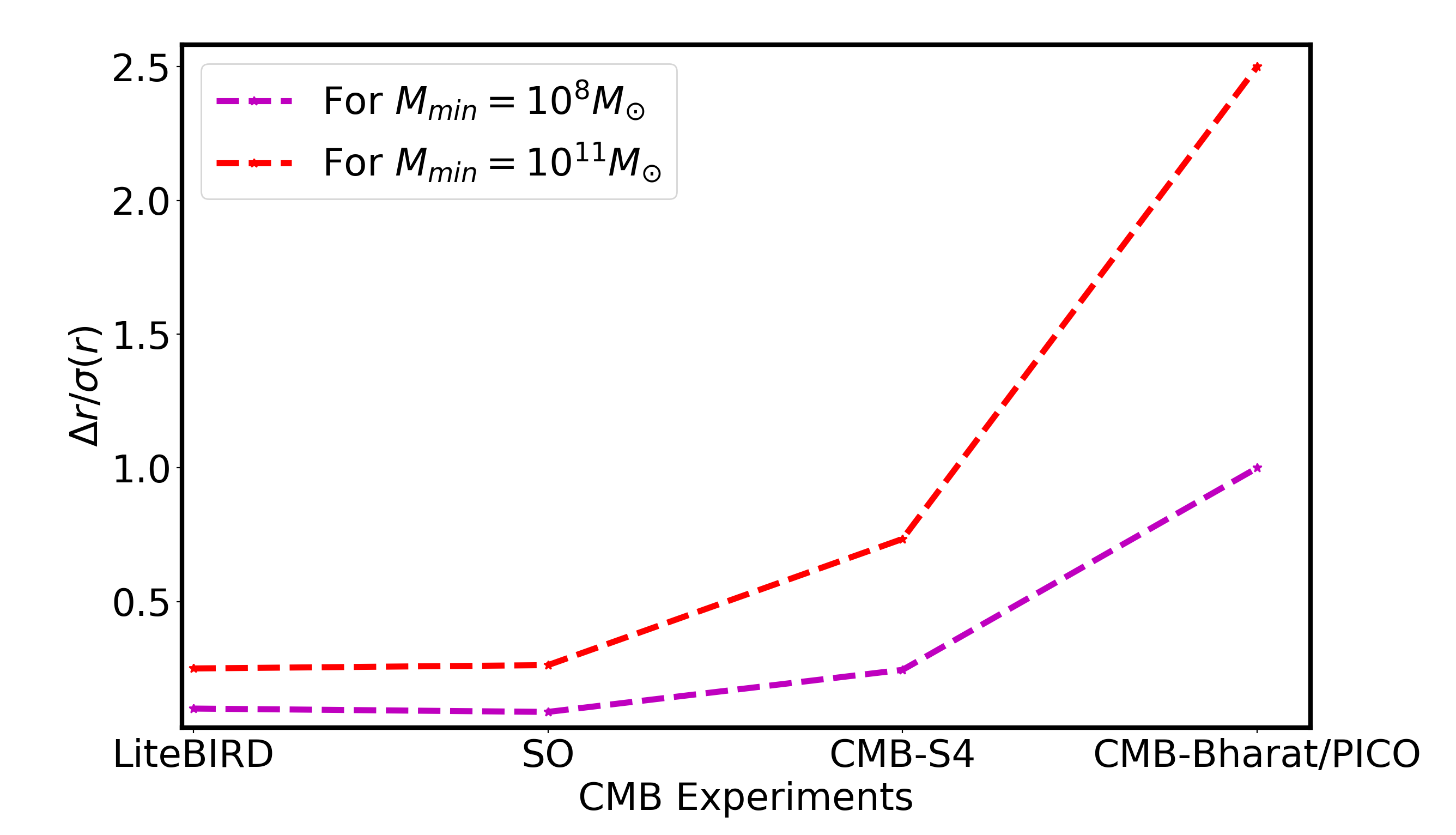}
\caption{ {The ratio of the bias $\Delta r = \hat r- r_{true}$ with respect to the $\sigma(r)$ for different CMB experiments due to two patchy reionization scenarios, assuming $r_{true} =10^{-3}$. We have taken $\sigma(r)= 10^{-3}$ (for LiteBIRD) and $\sigma(r)= 10^{-4}$ (for CMB-Bharat and PICO) with $l_{min}=2$. For the ground-based CMB experiments such as SO and CMB-S4, we have taken the same noise as mentioned in Sec. \ref{rcont} with $l_{min}= 50$.}}
\label{cmbmissions}
\end{figure} 

From the simulations considered in this analysis (which agree with several observational constraints on the reionization history), we find that in models with $r=10^{-3}$, one can get a maximum bias of about $30\%$ in the value of $r$ if $l<50$ are not accessible from the ground-based CMB experiments. If the low $l$ values are accessible from space-based missions, then the bias can be reduced to $25\%$ for the halo mass of $M_{min}= 10^{11}$ $M_\odot$. For the lowest halo mass $M_{min}= 10^8 M_\odot$ considered in this analysis, we find the bias in the value of $r$ to be about $10\%$. We show the variation of the bias with $M_{min}$ in Fig. \ref{rbias} for different choices of $l_{min}$. The bias in the value of $r$ found from our simulations are less than $0.5$--$\sigma$ for the CMB missions such as Simons Observatory \citep{Ade:2018sbj} and LiteBIRD \citep{Matsumura:2013aja} and hence are not going to contaminate the value of primordial gravitational wave signal severely. However,  {for the proposed CMB experiments with $\sigma(r) \leq 5 \times 10^{-4}$ such as CMB-S4 \citep{2016arXiv161002743A}, CMB-Bharat \footnote{\url{http://cmb-bharat.in/}} and PICO \citep{2018SPIE10698E..46Y,Hanany:2019lle}, the bias in the value of $r$ can be greater than 1-$\sigma$ as shown in Figure \ref{cmbmissions}. The amount of bias varies with the model of reionization and may be difficult to estimate this bias robustly from CMB polarization data alone.  
 
However, there are other cosmological observables such as HeII reionization, intergalactic Medium (IGM) temperature from the Lyman-$\alpha$ forest and $21$ cm power spectrum which can make independent measurements of the reionization history.  Scenarios of reionization driven by AGN will reionize HeII at higher redshifts and hence would lead to a very different thermal history at $z \sim 3$ as compared to the conventional reionization models. As a result, the models of reionization driven by $10^{11} \, M_\odot$ can be constrained from HeII reionization \citep{2011MNRAS.410.1096B} and a recent study has already imposed constraints on such scenarios \citep{Mitra:2016olz}. The power spectrum of the $21$ cm signal fluctuations is also expected to be larger for AGN driven reionization models \citep{Kulkarni:2017qwu} and as a result, this scenario can also be constrained from future $21$ cm data. So by jointly using the data of HeII reionization and $21$ cm power spectrum, we can obtain an upper limit on the strength of patchy reionization signal in CMB B-mode polarization. Patchy reionization also generates kinetic Sunyaev-Zeldovich  (kSZ) signal and we will explore the feasibility of measuring the imprints of patchy reionization by jointly using both kSZ as well as B-mode polarization signal in future work.

\section*{Acknowledgement}
S.M. would like to thank Nick Battaglia, Neal Dalal, Eiichiro Komatsu, Joseph Silk, David Spergel, Maxime Trebitsch and Benjamin D. Wandelt for useful discussions. S. M. would also like to thank Shaul Hanany and Lyman Page for useful comments on the draft.  The work of S.M. is supported by Simons Foundation and the Labex ILP (reference ANR-10-LABX-63) part of the Idex SUPER, and received financial state aid managed by the Agence Nationale de la Recherche, as part of the programme Investissements d'avenir under the reference ANR-11-IDEX-0004-02. The computational analyses of S. M. are performed in the Rusty cluster of the Flatiron Institute, Simons Foundation.  T.R.C. acknowledges support from the Associateship Scheme of ICTP, Trieste. While our paper was in preparation, we learned about a similar work (in preparation) by Girish Kulkarni, Daan Meerburg and Anirban Roy. We would like to thank them for pointing us to their work and for having useful discussions.

\bibliography{main}
\label{lastpage}

\end{document}